# High quality electron beam generation in a proton-driven hollow plasma wakefield accelerator

**Authors:**
Y. Li[1,2,a)], G. Xia[1,2], K. V. Lotov[3,4], A. P. Sosedkin[3,4], K. Hanahoe[1,2], O. Mete-Apsimon[1,2]

**Affiliations:**
[1]School of Physics and Astronomy, University of Manchester, Manchester M13 9PL, UK.
[2]Cockcroft Institute, Warrington WA4 4AD, UK.
[3]Budker Institute of Nuclear Physics, Novosibirsk 630090, Russia.
[4]Novosibirsk State University, Novosibirsk 630090, Russia.

Simulations of proton-driven plasma wakefield accelerators have demonstrated substantially higher accelerating gradients compared to conventional accelerators and the viability of accelerating electrons to the energy frontier in a single plasma stage. However, due to the strong intrinsic transverse fields varying both radially and in time, the witness beam quality is still far from suitable for practical application in future colliders. Here we demonstrate efficient acceleration of electrons in proton-driven wakefields in a hollow plasma channel. In this regime, the witness bunch is positioned in the region with a strong accelerating field, free from plasma electrons and ions. We show that the electron beam carrying the charge of about 10% of 1 TeV proton driver charge can be accelerated to 0.6 TeV with preserved normalized emittance in a single channel of 700 m. This high quality and high charge beam may pave the way for the development of future plasma-based energy frontier colliders.

---

a) Corresponding author. yangmei.li@manchester.ac.uk

## I. INTRODUCTION

Particle beam driven plasma wakefield accelerators (PWFA) have achieved enormous progress in recent decades [1,2] and have experimentally demonstrated electron-driven acceleration up to 85 GeV [3]. The achievable energy gain in a single stage is, however, limited by the transformer ratio [4] and thereby by the energy of the drive beam. Combining multiple acceleration stages [5,6] can, in principle, overcome the transformer ratio limit, but electron-driven PWFA requires tens to hundreds of stages for energy frontier collider applications, which is technically challenging [7-11]. The need for multiple stages inherently arises from the limited energy content of available drivers [12]. Laser driven wakefield accelerators (LWFA) are hindered by the same disadvantage. In spite of rapid growth of the laser peak power [13], single stage acceleration energy hasn't advanced that much in the last decade, going up from 1 GeV reached with 40 TW laser in 2006 [14] to 4.2 GeV record held from 2014 on [15]. Conservative designs of LWFA-based colliders thus also rely on staging [16-18]. On the other hand, high energy proton beams are the only available drivers capable of accelerating particles to TeV energy level in a single plasma stage in the near future.

In simulations, proton-driven plasma accelerators have already demonstrated electron acceleration up to TeV-level [19-21]. However, unlike the electron driver, positively charged driver attracts plasma electrons instead of fully expelling them away and forming an electron-free bubble. The resulting transverse wakefield is radially nonlinear and varies both in time and along the witness bunch. The variations of radial forces degrade the quality of the accelerated beam. In addition, multiple Coulomb scattering on plasma ions further deteriorates the beam emittance [22,23].

Hollow plasma-based acceleration is free from the aforementioned issues [24-34]. The hollow channel provides guiding for a laser or particle beam and forms an ion-free region with near zero, or weak and linear transverse field, which benefits focusing and helps to preserve the witness bunch quality. Recently, an encouraging experiment at FACET of SLAC National Accelerator Laboratory has demonstrated an 8 cm long hollow plasma channel with a radius of 250 μm, in which the measured decelerating gradient on the driving positron beam has reached 230 MeV/m [24]. This opens up prospects for practical applications of hollow plasmas in plasma wakefield acceleration.

Earlier theoretical studies of hollow channels were mostly focused on low amplitude waves and linear plasma responses [18,26-33] and on laser [18,26-29] or electron [30] drivers that push plasma electrons aside rather than pull them into the channel. Studies of positively charged drivers are fewer in number [32-36], but a good positron acceleration regime was found for this configuration [36]. In this regime, the plasma response is strongly nonlinear, and a wide area devoid of both plasma electrons and ions exists in the channel, in which the accelerating field approaches the wave-breaking limit. In simulations, the witness positron beam not only obtained a high energy, but also preserved its normalized emittance and reached a final energy spread of 1.5%. The new regime opens a path to accelerating a considerable amount of positrons to TeV-range energies with low emittance and energy spread.

However, future colliders also need high-energy and high-quality electrons. In this article, we propose and investigate the regime of electron acceleration in hollow channels, which is complementary to that of Ref. [36]. Similarly to positron acceleration in Ref. [36], the wakefield is driven by a short proton bunch in a hollow channel, and the witness beam can simultaneously have a high charge and experience a strong accelerating field. The latter features enable high energy transfer efficiencies, approaching those in the strong blowout regime in uniform plasmas driven by electrons [37]. The discovered similarity between electron and positron acceleration in hollow channels is not typical for nonlinear wakefields and even surprising. It contrasts with the uniform plasma case, for which the acceleration conditions are strongly charge-dependent [38]. For electron drivers in hollow channels, no equally good regimes have been found yet.

In Sec. II, we introduce the basic concept of the proposed scheme and then determine parameters for the particle-in-cell (PIC) simulations. Then we elucidate the transverse dynamics of the driving and witness bunches in the hollow plasma and the longitudinal acceleration of the witness bunch in Sec. III, where the conservation of low witness normalized emittance is confirmed. In Sec. IV we discuss the dependence of the energy gain and energy spread on the channel radius and plasma density and promote some approaches to further reduce the energy spread. After discussing the possibility of the instabilities, we give the conclusions in Sec. V.

## II. SCHEMATIC OF THE PROPOSED SCHEME AND PARAMETERS FOR PIC SIMULATIONS

In the uniform plasma, it is nearly impossible for the proton driver to create an electron-free blow-out area like the electron driver does, as the protons pull in the plasma electrons towards the propagation axis. The strongly nonlinear interaction might create a rarefied region of plasma electrons [2,19,20] but it requires the proton driver to be short enough to resonantly excite the plasma electron waves. The proposed proton-driven hollow plasma scheme, however, has the potential to realize an acceleration region devoid of plasma electrons. A simple schematic of this concept is illustrated in Fig. 1. The plasma electrons and ions are initially located outside a vacuum channel. When the proton bunch propagates through the hollow channel, it attracts plasma electrons into the channel and drives the wave, in which a considerable amount of electrons oscillates between the channel and the plasma. At certain wave phases, there are no electrons near the channel axis, but the longitudinal electric field is present. These regions are ideal for accelerating the witness beam [39], as the absence of both plasma electrons and ions results in no transverse wakefields and, thereby, conservation of the witness emittance. In the following, we demonstrate the efficient acceleration regime with simulations and discuss which of its specific features are responsible for particular advantages.

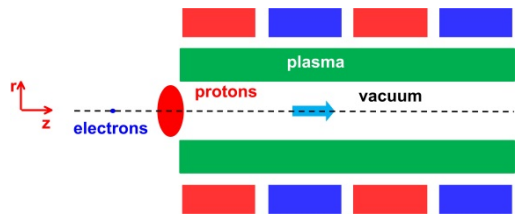

FIG. 1. Schematic of the concept of the proposed proton driven electron acceleration in a hollow plasma channel. The red and blue rectangles denote the quadrupole magnets with alternating polarity.

We use 2D axisymmetric quasi-static PIC code LCODE [40,41] to conduct hundreds of meters long simulations owing to its high computational efficiency. In the simulations, both the beam and the plasma are modeled by fully relativistic macro-particles. The simulation window using the coordinate system ($r, \theta, z-ct$) travels with the speed of light $c$. The window width is large enough (3 mm) to extend to the zero-field area, so the boundary conditions have no effect on the solution. The radial and axial grid sizes are $0.05c/\omega_p \approx 12$ μm. The time step is $8.404\omega_p^{-1} \approx 6.7$ ps, corresponding to the length of 2 mm.

The parameter set of the simulated case is detailed in the Table 1. Most importantly, the driver radius ($\sigma_{rd} = 350$ μm) is equal to the channel radius $r_c$, the driver length ($\sigma_{zd} = 150$ μm) is slightly shorter than the skin-depth of the outer plasma ($k_p^{-1} = 238$ μm), the initial witness radius ($\sigma_{rw} = 10$ μm) is much smaller than the channel radius, and the driver charge is approximately 20 times higher than the charge of electrons in volume $k_p^{-3}$ of the unperturbed plasma. The witness charge amounts to 8.7% of the driver charge. Driver population ($1.15\times10^{11}$ protons) and energy ($W_{d0} = 1$ TeV) are intentionally chosen close to Refs. [19], [20] and [36] to facilitate comparison.

The channel is surrounded by external quadrupole magnets, which define the axis of the system and keep the beams focused. The external quadrupoles were first proposed in Ref. [7] to precisely align the trajectory of the driver and to prevent the emittance driven erosion of the driver head. For short proton drivers that propagate hundreds of meters, the quadrupole focusing becomes a must [19]. In the hollow channel, not only the driver head but the whole driver and the witness are guided by quadrupoles, as there is no plasma focusing in the channel. According to Ref. [19], the quadrupole focusing works properly (i.e., the radial oscillations of the particles are insignificant) when the focusing period of quadrupoles is much shorter than the period of transverse particle oscillations, that is,

$$L_q \ll 2\pi\sqrt{\frac{W}{eS}} \qquad (1)$$

where $L_q$ is the quadrupole period, $W$ is the beam energy and $S$ is the magnetic field gradient. For an achievable $S$ of 0.5 T/mm and initial driver energy of $W$, the upper limit of $L_q$ for the driver is 16 m. As the witness bunch has a lower initial energy, the corresponding upper limit is smaller than the one for the driver. Since the quadrupole magnets are installed along the whole plasma channel, their period should satisfy the smaller upper limit, i.e. the one for the witness bunch. In addition, the simulation indicates that $L_q$ should be larger than 0.9 m in order to guarantee that over 98% of protons survive after 700 m propagation in the hollow plasma. We choose $S = 0.5$ T/mm and $L_q = 0.9$ m as the optimal quadrupole parameters which require a minimum quadrupole strength to sufficiently focus the driver. Based on the Eq. (1), the initial energy of the witness bunch must significantly exceed 3.1 GeV. 10 GeV is chosen to keep radial oscillations of the witness bunch small. Eq. (1) also implies that the increase of the witness energy during the acceleration will gradually alleviate the limitation to the quadrupole period.

**TABLE 1:** Parameters for simulation

| Parameters | Values | Units |
|---|---|---|
| **Initial driving proton beam:** | | |
| Proton population | $1.15\times10^{11}$ | |
| Initial energy, $W_{d0}$ | 1 | TeV |
| Energy spread | 10% | |
| Longitudinal bunch length, $\sigma_{zd}$ | 150 | µm |
| Beam radius, $\sigma_{rd}$ | 350 | µm |
| Angular spread | $3\times10^{-5}$ | |
| **Initial witness electron beam:** | | |
| Electron population | $1.0\times10^{10}$ | |
| Initial energy, $W$ | 10 | GeV |
| Energy spread, $\delta W/W$ | 1% | |
| Longitudinal bunch length, $\sigma_{zw}$ | 15 | µm |
| Beam radius, $\sigma_{rw}$ | 10 | µm |
| Angular spread | $1\times10^{-5}$ | |
| **Unperturbed hollow plasma:** | | |
| Plasma density, $n_p$ | $5\times10^{14}$ | cm$^{-3}$ |
| Hollow radius, $r_c$ | 350 | µm |
| Simulated length, $L$ | 700 | m |
| **External quadrupole magnet:** | | |
| Magnetic field gradient, $S$ | 0.5 | T/mm |
| Quadrupole period, $L_q$ | 0.9 | m |

### III. BEAM DYNAMICS IN THE HOLLOW PLASMA CHANNEL

The spatial structure of the focusing force in the efficient regime [Fig. 2] correlates with the density of plasma electrons. The field structure in general resembles that of the bubble or blowout regimes in a uniform plasma, but here the driver and witness reside in different bubbles. In the electron-free regions, the radial wakefield is exactly zero. Wherever plasma electrons enter the channel, the force focuses protons and defocuses electrons (blue areas in Fig. 2(a)). This requires the witness radius to be small. The radial structure of the focusing force on the driver is similar to that of a sharp reflecting wall (the green line in Fig. 2(b)), which is beneficial in two aspects. First, the driver emittance does not blow up as the driver comes into the radial equilibrium [42,43]. Second, as the energy $W_d$ of driving particles decreases, the amplitude of their radial (betatron) oscillations remains constant and does not increase as $W_d^{-1/4}$, as it does in uniform plasmas [39,44]. Consequently, driver depletion in energy does not result in driver widening and wakefield reduction.

The strong defocusing outside the witness bunch region (marked by the red dashed line in Fig. 2(a)) makes external quadrupoles a necessary part of the concept, as the quadrupoles control the witness radius. Note that the external magnetic fields of the quadrupoles are not included in the transverse fields in Fig. 2. The quadrupoles also protect the driver head from emittance-driven erosion that would otherwise shorten the acceleration distance [20], as the plasma focusing is weak there (the blue line in Fig. 2(b)).

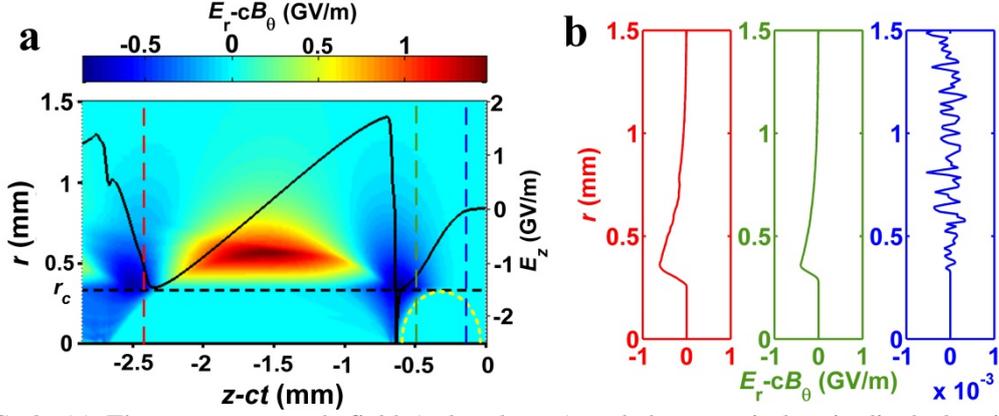

FIG. 2. (a) The transverse wakefield (colored map) and the on-axis longitudinal electric field (black line). The yellow dashed curve shows the driver location, and the black dashed line marks the channel boundary. The red dashed line is at the midpoint of the witness bunch. (b) The transverse wakefields at three longitudinal positions marked in (a) with vertical dashed lines of the corresponding colors. The simulation window travels to the right at the speed of light $c$.

The witness bunch initially resides slightly behind the accelerating field maximum, so that the bunch head experiences a stronger field. This trick lowers the energy spread, since the field slope at the witness location changes its sign as the driver depletes. Due to strong nonlinearity of the plasma response, the transformer ratio is about unity. The ratio of the average accelerating field acting on the witness to the average decelerating field acting on the driver is even higher.

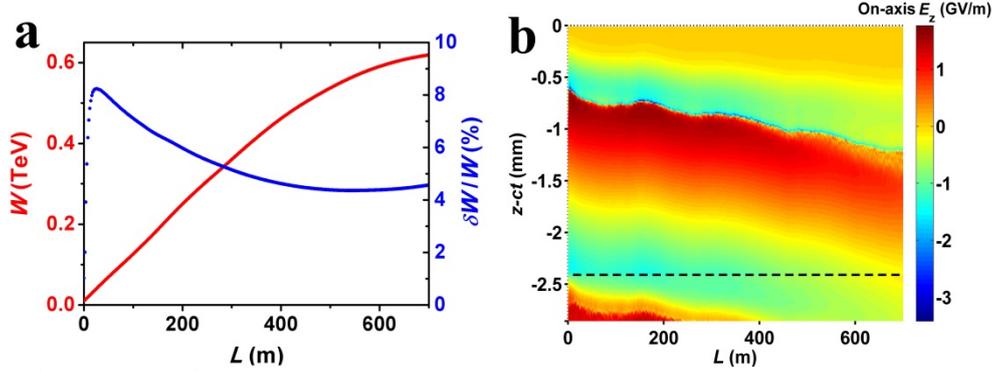

FIG. 3. Dependence of the mean witness energy and the relative energy spread (a) and the on-axis accelerating field (b) on the propagation distance. The black dashed line in (b) marks the longitudinal midpoint of the witness bunch.

The final energy gain of the witness (0.62 TeV ≈ 0.6 $W_{d0}$) and the required acceleration length (700 m) are comparable to the best results of Refs. [20] and [36] [Fig. 3(a)]. The beam energy increases linearly for the initial 400 m, afterwards the increase of the bunch energy slows down. Still, the average accelerating gradient exceeds 1.0 GeV/m over the first 600 m. The decrease of the acceleration rate mainly comes from the backward shift of the field pattern as a whole [Fig. 3(b)], which is caused by the relatively low relativistic factor of the driver and is unavoidable for driver energies of about 1 TeV or below [20]. However, since the witness initially resides at the rear of the second bubble, the acceleration continues until the wave phase shifts back by almost a half of the period [Fig. 4(a)] At this point, the shape of the drive bunch has changed significantly [Fig. 4(b)]. Nevertheless, since driving protons cannot escape from the channel, the driver still excites strong wakefields. The final energy spread of the witness in this particular case is 4.6% [Fig. 3(a)]. Note that here no optimization of the witness shape has been done in order to reduce the energy spread.

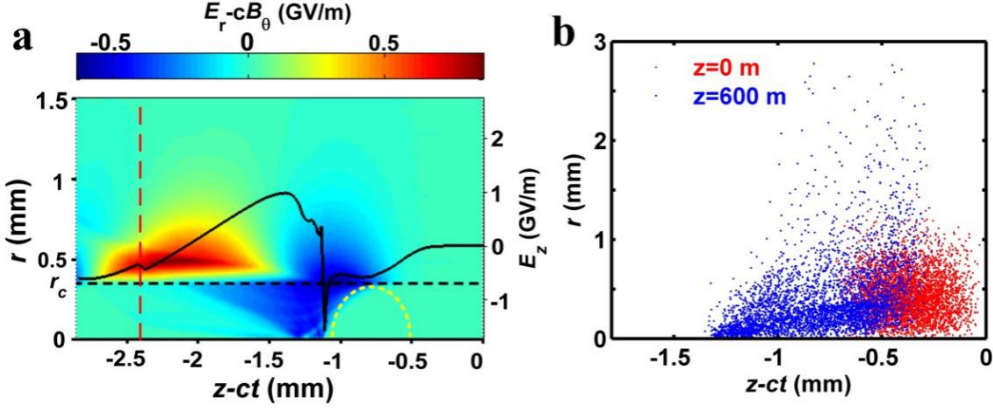

FIG. 4. (a) The corresponding transverse wakefield and the on-axis longitudinal electric field at $z = 600$ m in comparison with Fig. 2a. (b) Snapshots of the proton bunch before (red) and after (blue) the initial 600 m.

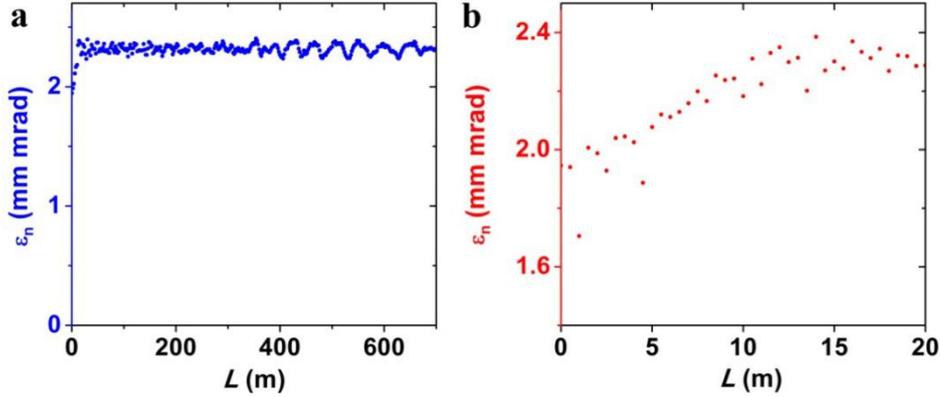

FIG. 5. (a) The normalized emittance of witness electrons over the whole acceleration process. (b) A close-up of the emittance growth in the first 20 m.

As expected, the normalized emittance of the witness bunch is preserved at the level of 2.4 mm mrad [Fig. 5(a)]. To our knowledge, this is the first time the emittance preservation at this small level has been directly demonstrated in simulations of plasma wakefield acceleration over such long distances. This proves applicability of available PWFA simulation codes for studies of low-emittance beams. Apparently, if the initially loaded emittance is reduced, a smaller final emittance is obtainable owing to no blow-up arising from the transverse plasma wakefields. Fig. 5(b) indicates that the beam emittance increases from the initial 1.95 mm mrad to 2.4 mm mrad during the first 15 m. The emittance growth is caused by the quadrupoles, as there are no transverse plasma wakefields acting on the witness. The initial witness radius (10 μm) is larger than the equilibrium radius [20] (7.6 μm) calculated for our quadrupole strength. During the first few meters of acceleration, the beam energy spread quickly blows up [Fig. 3(a)] and causes frequency variation in betatron oscillations and phase mixing. The incoherence of particle oscillations results in the emittance growth, but after 15 m the beam matches with the focusing structure, and the normalized emittance levels off. We deliberately present the initially unmatched case here to demonstrate the scale of this effect.

## IV. DISCUSSION

The simulated case is typical for a wide range of the parameter space. For example, variations of the channel radius and the plasma density here [Fig. 6] do not result in substantial changes of the energy gain and energy spread. Thus this allows for some freedom in optimization for some parameters, like acceleration distance, beam-to-beam efficiency, etc.

It follows from the Maxwell's equations that the accelerating wakefield is constant across the plasma electron-free part of the channel. Consequently, the energy spread of the witness bunch is the correlated one that appears because of different energies of different longitudinal slices. This opens the possibility of minimizing the energy spread by tailoring the witness shape. As a first step, it is possible to adjust the witness length and charge so as to flatten the accelerating field due to beam loading effect [Fig. 7]. For this trick to work, the witness beam must be located in the region of the positive field gradient, where $\partial E_z/\partial z > 0$ [Fig. 7(a)], otherwise the beam loading would increase the energy spread [38]. With witness beams of a special shape [29-30,37,45-47], it is possible to further reduce the energy spread while keeping high charges. Note that optimization for the energy spread has a weak effect on the energy gain [Fig. 7(b)].

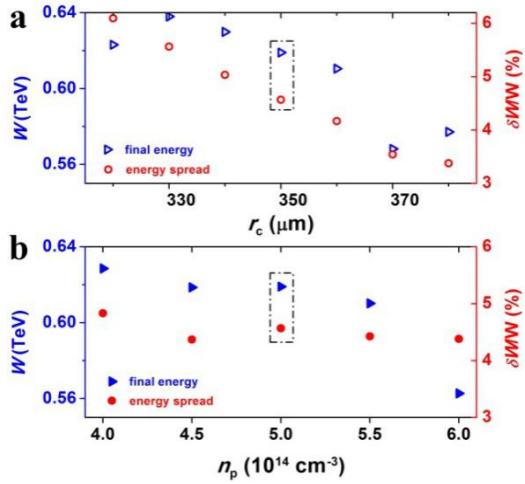

FIG. 6. Dependence of the energy gain and energy spread on the channel radius (a) and plasma density (b). The dash-dotted frames denote the simulated case. The acceleration length is individually optimized for each data set.

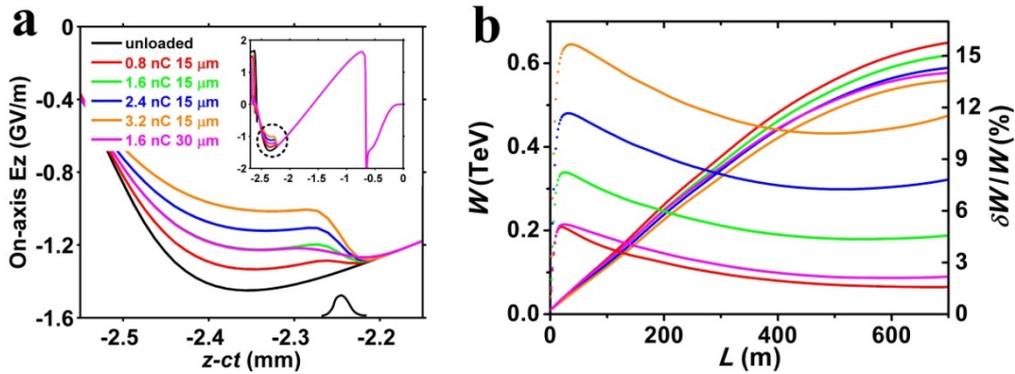

FIG. 7. (a) On-axis accelerating fields for unloaded and differently loaded cases with various charges and bunch lengths at z = 10 m. (b) Dependence of the mean energy and energy spread on the propagation distance for different witness bunches. The identically colored lines in (a) and (b) represent the same case.

The conducted two-dimensional axisymmetric simulations do not prove the stability of the proposed acceleration scheme, as full three-dimensional simulations are required for that. However, the experience gathered over decades of studying various beam-plasma instabilities suggests that this configuration is stable. Hosing-like instabilities [48-51] in which a transverse displacement of the beam couples with displacement of the surrounding plasma are a danger for long beams only. If the beam is about one plasma wavelength long or shorter, the hosing instability is suppressed [52,53], and experimental observations of the stable propagation of short beams confirm this [3,54]. Furthermore, large differences in oscillation frequencies suppress coupling of driver and witness transverse oscillations similarly to BNS damping in conventional accelerating structures [55,56]. Because of this difference, even trains of many bunches may stably propagate in plasma wakefield accelerators [57]. Resonances between transverse particle oscillations and sign-varying pushes from the quadrupoles are taken into account in our simulation model [7] and do not result in witness quality loss.

### V. CONCLUSIONS

In summary, we have new proposed a good regime for plasma wakefield acceleration of electrons in hollow channels and illustrated it with simulations. The proposed scheme solves the issue of beam normalized emittance degradation due to strong, radially- and time-varying focusing forces in a uniform plasma driven by a positively charged drive beam. In this regime, the witness bunch resides in a region with a strong accelerating field and an absence of plasma electrons. The region preserves its shape and location up to driver depletion, thus providing a high acceleration efficiency and an average transformer ratio of about unity. The witness beam takes advantage of the hollow channel; it experiences a radially uniform accelerating field and no transverse plasma fields. Therefore, the normalized beam emittance is conserved during acceleration, and the energy gain depends only on the longitudinal position along the bunch, thus enabling minimization of the energy spread by tailoring the bunch shape. The quadrupoles used to confine the driver from emittance-driven widening also play an important role in witness confinement. They keep the witness from entering the defocusing region that appears at large radii as the driver depletes. The resultant high quality electron beam with our proposed scheme may find applications in the next generation high energy frontier colliders.

## Acknowledgements

This work was supported by the President's Doctoral Scholarship Award of University of Manchester, the Cockcroft Institute core grant and STFC. The results mentioned in Chapter II were obtained under the support of the Russian Science Foundation (project no. 14-50-00080). The authors greatly appreciate the computing time from the clusters at the University of Manchester.